# Quantifying a qualitative framework of patients' perceptions, attitudes and behavior relevant to oral health related quality of life

Angelo M Passalacqua[1], Stephen Dunne[1], Timothy J Newton[1], Nairn HF Wilson[1], Ana NA Donaldson[1,2]

[1]King's College London Dental Institute, Biostatistics Unit; [2]Stony Brook University; Department of Applied Mathematics & Statistics

**Abstract**

Background: In a qualitative approach, Gregory et al (Gregory, Gibson, & Robinson, 2005) proposed a battery of items grouped in seven dimensions to reflect attitudes and behavior that are relevant to patients' ratings of their own oral health related quality of life (OHRQOL). The seven dimensions were: 1. *Normative* perception of own oral health relative to an average person; 2. *Attribution of Control* of oral health to self (*internal*) or to others (*external*) or to *Values-Importance of having good oral health,* or to *Adherence* to dentist advice; 3. *Trust in dentistry* (dentist and dental products); 4. *Accessibility/Availability* to good dental service for the patient and family; 5. Acceptance of dentistry as a *Commodity* product; 6. *Authenticity* preference over artificial beauty; and 7. *Character* Bias in the judgement of other people's oral health.

Aims: In this study we quantify the dimensions of Gregory's relevance framework and their influence on the change in OHRQOL in the context of an instantly-impacting intervention (tooth extraction) and the postoperative recovery period.

Methods: Ethical approval was obtained from the Research Ethics Committee at Guy's Hospital (Registration No. 08/H0804/10). 149 patients participated in the study. OHRQOL, the main outcome, was measured before, and twice after a tooth extraction, using the OHIP-14 instrument. In addition to the socio-demographic variables (age, gender, ethnicity, education level), the prognostic factors considered were: oral health related knowledge, behaviours and dental anxiety, as well as the seven dimensions of relevance to oral health proposed by Gregory *et al.* Statistical methods of analysis consisted of generalised estimating equations (GEE).

Results: Patient's trust in dental products, Normative perception of own oral health, preference for natural teeth, character bias in judgement and control by adherence to dentist's instructions, were all found to be significant factors in the longitudinal change of the OHIP-14. Borderline significance was found in terms of dental anxiety and symptoms. None of the socio-demographic characteristics were found significant.

Conclusion: Behaviour and attitudes, rather than socio-demographics and oral health related knowledge, influence change in OHRQOL. The dimensions of Gregory's relevance framework featured significantly in the models. Trusting that the dentist values the patient as a person and the importance the patient gives to having good oral health, are not found significant, yet adhering to dentist's advice has a beneficial effect on OHRQOL.

**Corresponding author:** Ana NA Donaldson; Email address: ana.donaldson@stonybrook.edu

**Introduction**

In a qualitative approach, Gregory et al (Gregory, Gibson, & Robinson, 2005) proposed a battery of items, grouped into seven dimensions, which directly influence the individual's response to the oral health related quality of life (OHRQOL) surveys. This battery of items conforms a construct termed "*relevance*" (of OHRQOL), describing the individual's own perception of their oral health and exposing their judgement of the system or others and other attitudes and behavior. Gregory's framework was proposed in a qualitative study but calls for the measurement of seven dimensions. The seven dimensions of Gregory's framework are described are:

- *Normative* perception or own oral health in relation to an average person.
- *Attribution of Control* to obtain or produce oral health to either: self (internal), others (external), values/importance of having good oral health or adhering to the advice of the dentist (adherence).
- *Trust* in dentistry: Trust in dentists (TID) and in dental products (TIDP).
- *Accessibility* and *Availability of* good dental services.
- Acceptance of dentistry as a *Commodity* product.
- *Authenticity/Naturalness* preference over artificial beauty.
- *Character* bias when judging teeth appearance in others.

In this study we quantify the dimensions of Gregory's relevance framework and assess their influence on the change in OHRQOL in the presence of an instantly-impacting intervention (tooth extraction) and in the postoperative follow-up. In addition to socio-economic characteristics (age, gender, ethnicity and education level), our models adjust for *oral health related knowledge, behavior and anxiety,* as potential confounders.

**Patients and Materials**

This study was observational on 149 tooth extraction patients that participated in the study. Ethical approval was obtained from the Research Ethics Committee at Guy's Hospital (Registration No. 08/H0804/10) for the collection of data and use of personal information. Patients had to be: age 18 years and older, without a mental illness, able to speak and read English, not undergoing treatment, and had a tooth extraction scheduled in 24 hours at the earliest for other than emergency reasons. Patients who were approached were given at least 24 hours to decide to join the study. Surveys were administered to the patients prior to treatment, two weeks after treatment and four weeks after treatment. The first follow-up was conducted two weeks after the extraction so that sensitivity to change could be evaluated in terms of the dramatic change expected after the impact intervention. The second follow-up took place 2 weeks later during which time no other impact intervention would have taken place.

**Outcome measure**

OHRQOL, was measured by the 14-item Oral Health Impact Profile (OHIP-14), a self-weighted survey (Locker et al. 2007). This instrument is based on the same dimensions of (general) health from the World

Health Organisation. Its reliability and validity have been established (Slade, 1997) and it has been used in past longitudinal studies (Allen & McMillan, 2003; Locker et al., 2004). The OHIP-14 elicits information about the importance each dimension has for the respondent. This scheme covers both, "*needs satisfaction*" (Higgs et al., 2003) and "*cultural context*" (Hofstede, 1984) aspects. The fact that disease prevalence varies between countries (Blum *et al.*, 2003) and oral diseases vary significantly between various communities suggest that environmental (culture and economic) factors play a significant role in oral health (Al Shamrany, 2006; Allen, 2003). A study comparing the unweighted Oral Health Impact Profile instrument in two different countries, Germany and England, found eight principal dimensions in the German sample and only seven in the English sample, suggesting that not even between two developed western countries can the same QOL instrument be used without avoiding culture bias. For this reason, we adopted the OHIP-14, including the self-assessment of the importance given to each dimension.

The weighted OHIP-14 comprises 14 questions, which are grouped in seven dimensions. Each question has two parts: the first part measures the impact of a particular problem on oral health and the second part measures the magnitude (weight) by which the impact bothered the patient. The two parts are intended to be answered together to determine a complete measure of impact. The OHIP total score is obtained by first multiplying these two parts for each question and then adding each product together to reach a final score. An OHIP score of zero means perfect oral health-related quality of life; the OHIP score increases with the worsening of a person's OHRQOL, to a maximum value of 168. A log-transformation is used to correct the skewedness often presented by the OHIP score.

**Potential prognostic factors for longitudinal change in the OHIP-14**

In addition to the socio-demographic variables (age, gender, ethnicity, education level), indicators related to the following factors were recorded:

- Oral health related knowledge, attitudes and behaviour,
- Self-perceived status of the patient's oral health,
- Oral health related symptoms,
- Dental anxiety and
- Dimensions of Gregory's relevance framework to oral health (Gregory *et al.*, 2005): normative, attribution of control, trust in dentistry, good dental services accessibility and availability, acceptance of dentistry as a commodity, preference for authenticity rather than artificial beauty, and character bias.

Knowledge of oral health issues is originally coded in five items relating to the benefits of: brushing teeth, flossing, using fluoridated toothpaste, and the dangers of eating too many sweets and using tobacco. Because of heterogeneity of the effects of knowledge of the dangers of tobacco in relation to the other items, we combine the first four and analyze this item separately. The validity of this construct has been explored (Zhu, Petersen, Wang, Bian & Zhang, 2003).

The dental anxiety dimension was originally coded in terms of five different scenarios (having a treatment the following day, sitting in the waiting room, about to have teeth scaled and polished, about to have a tooth prepared, about to have a local anaesthetic injection). In terms of homogenous effects, we combined the first three items together (into an item called *dental anxiety low threshold*) and the last three items together (into a subdimension called *dental anxiety stimulus driven*).

**Statistical Analysis and power calculation**

We used generalised estimating equations (GEE) to assess the effect of the covariates on the OHIP-14. Multivariate regressions were fitted in a stepwise manner to adjust for the effects of potential confounders and interactions. Multilevel modelling was used to take into account the repeated measures, clustered within patient. We based our power assessment on a previous study that used the Oral Health Impact Profile-49 in two groups of edentulous patients in the USA (Awad, 2000). In this reference study, group mean changes of 35 (SD=32) and 9 (SD=39) were found in two groups of 54 and 48 patients, respectively. This corresponded to a mean difference in difference of 25 (SD=36), or a standardised mean difference of 0.70. For our pairwise comparisons, our sample had 80% power to detect medium size effects of size 0.78 an above, a size similar to the effects found in the reference study. With 149 patients recruited into our study and an anticipated drop-out rate of 40% per follow-up, our observational study had 80% power, at the 5% significance level, to detect standardised longitudinal changes of size 0.40 and above, which, according to Cohen (1962), is a medium size effect.

**Results**

The baseline values of the socio-demographic patient characteristics (Table 1). The mean age of the 149 patients included in the study was 47 years (95% C.I. 44 to 49) and 47% were male. The ethnic distribution of the patients was: white (N=76; 55%), black (N=45; 32%) and Asian (N=16; 12%). Other covariates (Table 2) are shown, together with an assessment of the significance of the longitudinal changes of the covariates.

**Reliability of the newly operationalized dimensions of Gregory's framework**

Gregory's framework is qualitative and calls for the measurement of seven dimensions. At the time we conducted this study, only two had been completely explored previously: Attribution/Control (Borkowska, Watts & Weinman, 1998) and Accessibility/Availability of good dental services (Pechansky & Thomas, 1981) and one had been partially explored: the TIDP component of Trust in Dentistry (Anderson & Dedrick, 1990). Five dimensions were still unexplored and, in consequence, for this study, we generated an instrument to evaluate the remaining unexplored dimensions of Gregory et al. The intra-rater reliability tests yield kappa coefficients of 0.48 (SE= 0.15) for norm, 0.69 (SE=0.16) for Commodity, 0.48 (SE=0.15), and 0.53 (SE = 0.15) for Character bias. All these coefficients signal moderate to substantial agreement and the kappa coefficients for the three items defining trust in dental products are less but still in the region of fair agreement (Fleiss, 1971): 0.31 (SE=0.12), 0.17 (SE=0.11), 0.39 (SE=0.11). Summaries and the corresponding assessment of the significance of the longitudinal change of each of the seven Gregory's framework dimensions for our study sample are exhibited in Table 3.

**Responsiveness of the OHIP-14 measure**

Table 4 exhibits the summaries of the QOL outcome for each time of assessment and the assessment of the significance of its longitudinal change. There was a significant time trend in the mean OHIP-14 from baseline to the last follow-up. The mean log-OHIP-14 scores (standard deviation) for the baseline and the two follow up times were 3.33 (0.97), 2.53 (1.30), 2.03 (1.46) respectively. An analysis of variance with repeated measures signalled a significant change in this outcome across time (P=0.001). On further (post-anova) analysis, we found that this significant difference was related to the change at both follow up times. The change between baseline and two weeks was -0.80 (= 2.53-3.33). At the median OHIP-14 this corresponds to a reduction of 33 in the original OHIP-14 scale. At the lower and upper quartiles, this corresponds to reductions of -20.8 and -60 respectively. The mean change between two and four weeks was -0.50; at the median OHIP-14 this corresponds to a reduction of -17.5 in the original OHIP-14 scale. At the lower and upper quartiles, this corresponds to reductions of -11 and -31.7 respectively.

**Complete case analysis**

The generalized estimating equation model obtained by complete case analysis is presented in Table 5. A significant time effect (b = -0.63; 95% C.I. (-0.80,-0.46); p=0.000) was detected. At the median OHIP-14 this corresponds to a reduction of 12.6 in the original QOL OHIP-14 scale. At the lower and upper quartiles, this corresponds to reductions of 7.9 and 22.9 respectively. After adjusting for the time effect, the most important variables that had a significant effect on OHIP were anxiety, symptom and six (of seven) of Gregory's Relevance dimensions: normal, control, TIDP, naturalness, character and accessibility/availability. Knowledge of oral health issues was not statistically significant. The only covariates that varied with time were: naturalness, floss behaviour and symptom (bleeding gums) and satisfaction. Of these variables, only satisfaction was found to have had a significant interaction with time. Five of the seven dimensions in Gregory's framework were found to have a significant association with change in OHRQOL: *normative, locus of control (external and adherence)*, the trust subdimension TIDP, *character bias and naturalness* (Table 5). The trust subdimension *trust in dentists* was not found to be significant (p=0.81). The remaining dimensions, *acceptance of dentistry as a commodity* (p=0.50) and *accessibility/availability* of good dental care for the patient and family (P=0.10) were not found to have a significant effect.

**Interpretation of the model:**

- For *normative*, the mean log QOL OHIP-14 score (lnQOL) was 0.38 lower (indicating a better QOL) for people who believe their oral health is much worse or worse than people who believe their oral health is about the same or better (95% C.I. (-0.65, -0.11); P=0.01). At the median OHIP-14 this corresponds to a reduction of 12.5 in the original QOL OHIP-14 scale. At the lower and upper quartiles, this corresponds to reductions of 7.9 and 22.7 respectively.
- For *control-external*, the mean log QOL OHIP-14 score (lnQOL) was 0.11 lower (indicating a better QOL) for each unit increase in control (others) score (95% C.I. (-0.22, 0.01); P=0.07). At the median OHIP-14 this corresponds to a reduction of 2.8 in the original QOL OHIP-14 scale. At the lower and upper quartiles, this corresponds to reductions of 1.7 and 5.1 respectively.
- For *control-adherence*, the mean log QOL OHIP-14 score (lnQOL) was 0.25 higher (indicating a worse QOL) for people who agree or strongly agree with the phrase, "I find it difficult to do exactly what the dentist tells me to do", than people who strongly disagree or disagree (95% C.I. (-0.004, 0.05); P=0.05). At the median OHIP-14 this corresponds to a increase of 7.6 in the original QOL OHIP-14 scale. At the lower and upper quartiles, this corresponds to increases of 4.8 and 13.9 respectively.
- For *TIDP*, the mean log QOL OHIP-14 score (lnQOL) was 0.09 lower (indicating a better QOL) for each unit increase in the continuous scale of trust in dental products score (95% C.I. (-0.16 to -0.03); P=0.003). At the median OHIP-14 this corresponds to a reduction of 2.5 in the original QOL OHIP-14 scale. At the lower and upper quartiles, this corresponds to reductions of 1.6 and 4.6 respectively.
- For *Character bias*, the mean log QOL OHIP-14 score (lnQOL) was 0.37 higher (indicating a worse QOL) for people who would admire the healthy teeth of strangers in relation to people who are indifferent or believe those strangers are vein or insincere (95% C.I. (0.07, 0.67); P=0.02). At the median OHIP-14 this corresponds to an increase of 12 in the original QOL OHIP-14 scale. At the lower and upper quartiles, this corresponds to increases of 7.6 and 21.9 respectively.
- For *naturalness preference*, the mean log QOL OHIP-14 score (lnQOL) was 0.28 lower (indicating a better QOL) for people who would rather have less perfect (authentic) teeth than people who would rather have perfect (artificially improved) teeth (95% C.I. (-0.57, 0.01); P=0.06). At the median OHIP-14 this corresponds to a reduction of 8.7 in the original QOL OHIP-14 scale. At the lower and upper quartiles, this corresponds to reductions of 5.4 and 15.8 respectively.
- For *accessibility/availability of good dental services for patient and family*, the mean log QOL OHIP-14 score (lnQOL) was 0.02 lower (indicating a better QOL) for each unit increase in access score (95% C.I. (-0.05, 0.004); P=0.10). At the median OHIP-14 this corresponds to a reduction of 0.5 in the original QOL OHIP-14 scale. At the lower and upper quartiles, this corresponds to reductions of 0.3 and 0.9 respectively.

- For *dental anxiety low-threshold*, the mean log QOL OHIP-14 score (lnQOL) was 0.06 higher (indicating a worse QOL) for each unit increase in dental anxiety score (95% C.I. (-0.003, 0.12); P=0.06). At the median OHIP-14 this corresponds to an increase of 1.6 in the original QOL OHIP-14 scale. At the lower and upper quartiles, this corresponds to increases of 1.0 and 3.0 respectively.
- For *symptom*, the mean log QOL OHIP-14 score (lnQOL) was 0.29 higher (indicating a worse QOL) for people who sometimes, almost always or always bleed when they brush or floss compared to those who never or rarely bleed (95% C.I. (-0.02, 0.60); P=0.07). At the median OHIP-14 this corresponds to an increase of 9.0 in the original QOL OHIP-14 scale. At the lower and upper quartiles, this corresponds to increases of 5.7 and 16.4 respectively.

*Dental anxiety stimulus-driven* (p=0.16) and *flossing habits* (p=0.28) were not statistically significant. The locus of control variables were non significant for: internal (P=0.17) or values-importance (p=0.93). Knowledge of oral health issues did not have a significant effect on OHRQOL. The mean log QOL OHIP-14 score (lnQOL) is 0.15 lower (suggesting a better QOL) for people who scored high on knowledge level in relation to people who scored medium/low (95% C.I. (-0.45 to 0.16); P=0.34). At the median OHIP-14 this corresponds to a reduction of -4.3 in the original QOL OHIP-14 scale. At the lower and upper quartiles, this corresponds to reductions of -2.7 and -7.9 respectively.

Little data was found missing within returned surveys, however the study experienced about 40% attrition per time point. On logistic regression, the only independent variable that was associated with attrition was age; the older the patient, the less likelihood of attrition (OR=0.97, 95% C.I. (0.94,0.99);p=0.01). Missing values of outcome at baseline were independent of the satisfaction with own oral health variable, which was essentially a proxy measure for OHRQOL. Missing values for the outcome, OHIP-14, at any follow-up time was not dependent on previously observed values of the OHIP-14 score (p=0.51 for first follow-up and p=0.87 for the second follow-up) and were also independent of the level of satisfaction at any time (p=0.16 and p=0.58 respectively). This suggests that the missing data can be classified as MAR. Multiple imputation (Rubin, 1987) was conducted to account for potential bias introduced from missing data. Analyses from the multiple imputation process consistent with the complete case analysis.

**Discussion**

Behaviour and attitudes influence change in OHRQOL. With the exception of commodity, the dimensions of Gregory's relevance framework feature significantly in the models: Patient's trust in dental products, normative perception of own oral health, preference for natural teeth, character bias in judging the oral health of others and control by adherence to dentist's instructions, are all found to be significant factors in the longitudinal change of the OHIP-14. Trusting that the dentist values the patient as a person and the importance the patient gives to having good oral health are not found to significantly influence change in OHRQOL, yet a significant beneficial effect of adhering to dentist's advice is found. Borderline significance is found in terms of *dental anxiety low-threshold* and *symptom*s. None of the socio-demographic characteristics, neither the oral health related knowledge is found to significant influence change in OHRQOL.

Our approach to the evaluation of factors that influence change in oral health related quality of life for the sample of patients referred for tooth extraction leads us to some significant contributions to dental research methods.

One contribution is in relation to operationalising Gregory's framework for relevance to oral health. This was established by means of a qualitative study (Gregory et al, 2005). First, four single-item dimensions, related to Gregory's framework, were generated, tested and their reliability assessed. One such dimension was *commodity*. A series of measures to describe this dimension has been proposed (Birch & Ismail, 2002). The significance of the variables *naturalness* and TIDP in the final multivariate model, argues for the importance for future research on the properties (*validity* and *reliability*) of the proposed measure for these dimensions. Our study could well be the first of a series of studies to look at the impact of the seven dimensions on QOL and to test the new items' *reliability* and *validity* for future research.

Another contribution to oral health research relates to the measure of dental anxiety. Five items of the Modified Dental Anxiety Score (MDAS: Humphris et al., 1995) are factored in two dimensions. The three elements of being anxious, while only simply waiting for a dentist appointment the next day, or while simply in the waiting room or waiting to have teeth cleaned and polished are grouped into a dimension which we defined as "dental anxiety low threshold". The two items of being anxious while waiting for a tooth to be prepared, or waiting for an anaesthetic injection were grouped into a second dimension, which we defined as "dental anxiety stimulus driven". It was the first one, the dental anxiety low threshold, which showed to have a significant effect on QOL change, indicating a better characterisation of those patients that are really anxious, with a low-threshold for anxiety, perhaps over reacting to rather trivial dental interventions. This could be taken into account for future research, to present a theoretical model to improve definition and sensitivity of the anxiety scale.

A third contribution was the responsiveness of the oral-health related quality of life measure, the OHIP-14. Locker et al. studied responsiveness of the OHIP-14 in a group of 116 low-income elderly patients attending four municipally funded dental clinics. In their study, the responsiveness of the OHIP-14, a patient-based assessment, was assessed against another patient-based assessment of QOL, a global indicator of improvement between the two times. Nevertheless *Responsiveness* (or sensitivity to change) however is a property too often overlooked in studies that establish the validity of a new instrument and is not well established in OHRQOL instruments (Locker *et al.*, 2004). Our study lends an excellent ground to explore this property as it covers an impact intervention (the tooth extraction) that should bring change in the outcome for patients. To accomplish this, we assessed responsiveness of the OHIP-14 by assessing the significance of change under the scenario where change is present, brought about by the tooth extraction. The second follow-up may have captured the long-term effect of the intervention.

Our study has some limitations. First, we followed up patients for a maximum of fours weeks only. The time-frame of our study was sufficient to study the responsiveness of the OHIP-14, but was rather limited in terms of the ideal time span to allow the dimensions of relevance in Gregory's framework to manifest further. Further research, should aim for a longer follow-up and for larger sample size to allow the other dimensions of the relevance framework to show an effect in the change of OHRQOL. Second, although our patient population was not disease specific, it may have been specific in the sense that they had to have a tooth removed. Future work on the responsiveness of the OHIP could improve on this. Third, future research using structural equation modelling will allow mediating and moderating relationships to be fully explored. Finally, measurement or instrument error cannot be ruled out and, in this sense, the method of operationalisation could be made more rigorous; only one item representing a dimension was used in, for example, the *commodity* dimension. In this sense, our results for the Commodity and Accessibility/Availability dimensions could be deemed as inconclusive.

**Acknowledgements:** This paper was presented, by the first author, at the International Association of Dental Research Conference (IADR, Barcelona, 2010). The research was funded by an internal King's College London Dental Institute PhD studentship awarded to the first author. The authors are grateful to Nicholas Donaldson for his help with the typesetting of the tables and manuscript and for proof-reading several versions of this paper.

**Compliance with Ethical Standards**: This study was part of the PhD course of Dr Angelo Passalacqua (first author), in King's College London Dental Institute, with Professor Nora Donaldson (last author) as first supervisor. All authors declare they have no conflict of interest. Ethical approval was obtained from the Research Ethics Committee at Guy's Hospital (Registration No. 08/H0804/10) to collect data and use of personal information. Informed Consent was obtained from all individual participants included in this study.

## Appendix – Tables

**Table 1: Socio-demographics and baseline characteristics of the sample (n=149)**

|  | Mean (SD) / (missing) |
|---|---|
| **Age** | 46.7 (14) / (13) |
|  | n (%) |
| **Do you have dentures? (Y/N).** Yes: | 26 (17.4%) |
| **If you have dentures, do they feel loose in your mouth? (Y/N).** Yes: | 18 (12%) |
| **Have you had another tooth extraction? (Y/N).** Yes: | 9 (6%) |
| **Ethnicity** |  |
| White: | 76 (55%) |
| Black: | 45 (32%) |
| Asian: | 16 (12%) |
| **Education** |  |
| GCSE or equivalent: | 41 (30%) |
| A-levels or equivalent: | 19 (14%) |
| Degree (BSc, BA, or equivalent): | 37 (27%) |
| Advanced degree (MSc, MA, MBA or equivalent): | 11 (8%) |
| Research degree: | 1 (1%) |
| Other: | 26 (19%) |
| **Gender** Female: | 217 (53%) |
| **Relationship status** |  |
| Single: | 124 (31%) |
| In a relationship: | 93 (23%) |
| Married: | 124 (31%) |
| Separated: | 8 (2%) |
| Widowed: | 18 (4%) |
| Divorced: | 39 (10%) |

Table 2. Summaries and longitudinal changes of covariates

| (N=149) | Baseline | 2 weeks after | 4 weeks after |
|---|---|---|---|
| **Symptom** | | | |
| **Do your gums bleed when you brush or floss? (p=0.000)** | | | |
| Never | 16 (11%) | 12 (16%) | 3 (7%) |
| Rarely | 32 (23%) | 18 (25%) | 17 (39%) |
| Sometimes | 73 (51%) | 38 (52%) | 23 (52%) |
| Almost always/always | 21 (15%) | 5 (7%) | 1 (2%) |
| Missing | 7 | 76 | 105 |
| **Dental anxiety** | | | |
| **If you went to your dentist for treatment tomorrow, how would you feel? (p=0.08)** | | | |
| Not anxious | 39 (26%) | 28 (36%) | 15 (31%) |
| Slightly anxious | 54 (36%) | 25 (32%) | 24 (49%) |
| Fairly anxious | 30 (20%) | 14 (18%) | 5 (10%) |
| Very anxious/Extremely anxious | 25 (17%) | 11 (14%) | 5 (10%) |
| Missing | 1 | 71 | 100 |
| **If you were sitting in the waiting room (waiting for treatment), how would you feel? (p=0.06)** | | | |
| Not anxious | 32 (22%) | 28 (36%) | 15 (31%) |
| Slightly anxious | 57 (39%) | 24 (31%) | 21 (43%) |
| Fairly anxious | 27 (18%) | 11 (14%) | 8 (16%) |
| Very anxious/Extremely anxious | 32 (22%) | 15 (19%) | 5 (10%) |
| Missing | 1 | 71 | 100 |
| **If you were about to have a tooth drilled, how would you feel? (p=0.15)** | | | |
| Not anxious/Slightly anxious | 58 (39%) | 37 (47%) | 25 (51%) |
| Fairly anxious | 34 (23%) | 17 (22%) | 10 (20%) |
| Very anxious | 27 (18%) | 12 (15%) | 9 (18%) |
| Extremely anxious | 28 (19%) | 12 (15%) | 5 (10%) |
| Missing | 2 | 71 | 100 |
| **If you were about to have your teeth scaled and polished, how would you feel? (p=0.13)** | | | |
| Not anxious | 50 (34%) | 35 (45%) | 22 (45%) |
| Slightly anxious | 46 (32%) | 15 (19%) | 15 (31%) |
| Fairly anxious | 24 (16%) | 13 (17%) | 6 (12%) |
| Very anxious/Extremely anxious | 26 (18%) | 15 (19%) | 6 (12%) |
| Missing | 3 | 71 | 100 |
| **If you were about to have a local anaesthetic injection in your gum, above an upper back tooth, how would you feel? (p=0.15)** | | | |
| Not anxious | 20 (14%) | 14 (18%) | 7 (15%) |
| Slightly anxious | 41 (28%) | 25 (32%) | 18 (38%) |
| Fairly anxious | 38 (26%) | 17 (22%) | 13 (27%) |
| Very anxious | 21 (14%) | 9 (12%) | 6 (13%) |
| Extremely anxious | 28 (19%) | 13 (17%) | 4 (8%) |
| Missing | 1 | 71 | 101 |
| **If about to have a tooth drilled or local aesthetic injection in your gum, above an upper back tooth, how would you feel? (p=0.19)** | | | |
| Not anxious | 20 (14%) | 13 (17%) | 7 (15%) |
| Slightly anxious | 64 (44%) | 36 (46%) | 24 (50%) |
| Fairly anxious/Very anxious | 39 (27%) | 21 (27%) | 13 (27%) |
| Extremely anxious | 24 (16%) | 8 (10%) | 4 (8%) |
| Missing | 2 | 71 | 101 |
| **Regularity** | | | |
| **Do you only visit the dentist when something is wrong? (p=0.31)** | | | |
| No | 49 (33%) | 33 (43%) | 19 (39%) |
| Yes | 98 (67%) | 44 (57%) | 30 (61%) |
| Missing | 2 | 72 | 100 |
| **How long ago was your last dental visit to a dentist?** | | | |
| Within the last year | 89 (61%) | | |
| More than one year ago | 56 (39%) | | |
| Missing | 4 | | |

**Table 3. Summaries and longitudinal changes of the seven Gregory's Framework dimensions**

| (N=149) | Baseline | 2 weeks after | 4 weeks after |
|---|---|---|---|
| **Normative** | | | |
| **How do I perceive oral health relative to the average person (p=0.41)** | | | |
| Much worse/Worse | 36 (25%) | 12 (15%) | 11 (22%) |
| About the same/ Better/ Much better | 108 (75%) | 66 (85%) | 38 (78%) |
| Missing | 5 | 71 | 100 |
| **Locus of control** | | | |
| **Internal - I believe I can prevent gum disease. (p=0.83)** | | | |
| Strongly disagree/Disagree/Agree | 95 (65%) | 49 (63%) | 32 (64%) |
| Strongly agree | 51 (35%) | 29 (37%) | 18 (36%) |
| Missing | 3 | 71 | 99 |
| **External- Only the dentist can prevent gum disease. (p=0.30)** | | | |
| Strongly disagree | 34 (23%) | 23 (30%) | 12 (24%) |
| Disagree | 85 (58%) | 45 (58%) | 31 (63%) |
| Agree/Strongly agree | 27 (18%) | 9 (12%) | 6 (12%) |
| Missing | 3 | 72 | 100 |
| **Values- I believe that brushing and flossing are important and could possibly prevent gum disease and tooth loss, however I don't want to take the time today for something that may happen in 5 to 10 years from now. (p=0.50)** | | | |
| Strongly disagree | 40 (28%) | 25 (32%) | 15 (30%) |
| Disagree | 49 (34%) | 28 (35%) | 18 (36%) |
| Agree/Strongly agree | 55 (38%) | 26 (33%) | 17 (34%) |
| Missing | 5 | 70 | 99 |
| **Values- Good health is only of minor importance in being happy. (p=0.30)** | | | |
| Strongly disagree | 78 (53%) | 47 (59%) | 27 (54%) |
| Disagree | 40 (27%) | 24 (30%) | 20 (40%) |
| Agree/Strongly agree | 28 (19%) | 8 (10%) | 3 (6%) |
| Missing | 3 | 70 | 99 |
| **Values - There are many more things I care about than my health. (p=0.64)** | | | |
| Strongly disagree | 61 (42%) | 35 (44%) | 16 (32%) |
| Disagree | 49 (34%) | 24 (30%) | 24 (48%) |
| Agree/Strongly agree | 35 (24%) | 20 (25%) | 10 (20%) |
| Missing | 4 | 70 | 99 |
| **Adherence - It may be difficult for me to do exactly what the dentist told me to do. (p=0.23)** | | | |
| Strongly disagree | 28 (19%) | 10 (13%) | 6 (12%) |
| Disagree | 75 (52%) | 44 (56%) | 27 (54%) |
| Agree/Strongly agree | 42 (29%) | 24 (31%) | 17 (34%) |
| Missing | 145/4 | 78/71 | 50/99 |
| **Locus of Control Score (p=0.95)** | 13.14 (2.53) | 12.79 (2.4) | 12.90 (1.98) |
| Missing | 9 | 73 | 100 |
| **Trust in Dentistry** | | | |
| **Trust in dentists** | | | |
| **I doubt that my dentist really cares about me as a person. (p=0.06)** | | | |
| Strongly disagree/Disagree | 26 (18%) | 12 (16%) | 4 (8%) |
| Neutral | 88 (61%) | 38 (49%) | 31 (63%) |
| Agree/Strongly agree | 31 (21%) | 27 (35%) | 14 (29%) |
| Missing | 4 | 72 | 100 |
| **My dentist is usually considerate of my needs and puts them first. (p=0.13)** | | | |
| Strongly disagree/Disagree | 22 (15%) | 10 (13%) | 3 (6%) |
| Neutral/Agree/Strongly agree | 124 (85%) | 68 (87%) | 45 (94%) |
| Missing | 3 | 71 | 101 |
| **I trust my dentist so much that I always try to follow his/her advice. (p=0.07)** | | | |
| Strongly disagree/Disagree/Neutral | 71 (50%) | 32 (41%) | 17 (36%) |
| Agree/Strong agree | 72 (50%) | 47 (59%) | 30 (64%) |
| Missing | 6 | 70 | 102 |
| **If my dentist tells me something is so, then it must be true. (p=0.71)** | | | |
| Strongly disagree/Disagree | 20 (14%) | 8 (10%) | 6 (13%) |
| Neutral | 58 (41%) | 34 (44%) | 23 (49%) |
| Agree/Strongly agree | 64 (45%) | 36 (46%) | 18 (38%) |
| Missing | 7 | 71 | 102 |
| **I sometimes distrust my dentist's opinion and would like a second one. (p=0.12)** | | | |
| Strongly disagree/Disagree | 49 (34%) | 18 (23%) | 10 (21%) |
| Neutral | 62 (43%) | 39 (50%) | 26 (54%) |
| Agree/Strongly agree | 33 (23%) | 21 (27%) | 12 (25%) |
| Missing | 5 | 71 | 101 |
| **I trust my dentist's judgement about my dental care. (p=0.73)** | | | |
| Strongly agree/agree | 17 (12%) | 7 (9%) | 6 (13%) |
| Neutral | 33 (23%) | 23 (29%) | 11 (23%) |
| Agree | 75 (52%) | 36 (46%) | 26 (54%) |
| Strongly agree | 20 (14%) | 12 (15%) | 5 (10%) |
| Missing | 4 | 71 | 101 |
| **I feel my dentist does not do everything he/she should do for my dental care. (p=0.12)** | | | |
| Strongly disagree/Disagree | 39 (27%) | 14 (18%) | 6 (13%) |
| Neutral | 54 (38%) | 39 (50%) | 22 (46%) |
| Agree/Strongly agree | 50 (35%) | 25 (32%) | 20 (42%) |
| Missing | 6 | 71 | 101 |
| **I trust my dentist to put my dental needs above all other considerations when treating my dental problems. (p=0.46)** | | | |
| Strongly disagree/Disagree | 19 (13%) | 13 (16%) | 6 (13%) |
| Neutral | 53 (37%) | 26 (33%) | 14 (29%) |
| Agree/Strongly agree | 71 (50%) | 40 (51%) | 28 (58%) |
| Missing | 6 | 70 | 101 |
| **My dentist is a real expert when taking care of my dental problems. (p=0.09)** | | | |
| Strongly disagree/disagree/neutral | 75 (52%) | 39 (49%) | 18 (38%) |
| Agree/strongly agree | 68 (48%) | 40 (51%) | 30 (62%) |
| Missing | 6 | 70 | 101 |
| **I trust my dentist to tell me if a mistake was made about my treatment. (p=0.99)** | | | |
| Strongly disagree/disagree | 19 (13%) | 15 (19%) | 7 (15%) |
| Neutral | 53 (37%) | 21 (27%) | 15 (31%) |
| Agree | 50 (35%) | 34 (43%) | 22 (46%) |
| Strongly Agree | 20 (14%) | 9 (11%) | 4 (8%) |
| Missing | 7 | 70 | 101 |
| **I sometimes worry that my dentist may not keep the information we discuss totally private. (p=0.30)** | | | |
| Strongly disagree/Disagree | 17 (12%) | 11 (14%) | 4 (9%) |
| Neutral | 76 (53%) | 42 (53%) | 21 (45%) |
| Agree/Strongly agree | 51 (35%) | 26 (33%) | 22 (47%) |
| Missing | 144/5 | 79/70 | 47/102 |
| **Trust in dentists score (p=0.18)** | 3.4 (5.97) | 3.9(6.34) | 3.8 (5.36) |
| Missing | 13 | 73 | 104 |
| **Trust in dental products** | | | |
| **I trust dental products (tooth paste, tooth brush, floss, etc.) in general to be mostly safe. (p=0.69)** | | | |
| Strongly agree/Agree/Neutral | 34 (23%) | 13 (16%) | 11 (22%) |
| Agree | 88 (60%) | 54 (68%) | 34 (69%) |
| Strongly Agree | 24 (17%) | 12 (15%) | 4 (9%) |
| Missing | 3 | 70 | 100 |
| **I trust dental products (tooth paste, tooth brush, floss, etc.) in general to be mostly useful. (p=0.13)** | | | |
| Strongly disagree/disagree/neutral | 20 (14%) | 11 (14%) | 8 (16%) |
| Agree | 99 (68%) | 56 (71%) | 38 (78%) |
| Strongly agree | 26 (18%) | 12 (15%) | 3 (6%) |
| Missing | 4 | 70 | 100 |
| **I trust dental products (tooth paste, tooth brush, floss, etc.) in general to be mostly reliable. (p=0.96)** | | | |
| Strongly disagree/disagree/Neutral | 32 (22%) | 11 (14%) | 8 (16%) |
| Agree | 92 (63%) | 56 (71%) | 38 (78%) |
| Strongly Agree | 22 (15%) | 12 (15%) | 3 (6%) |
| Missing | 146/3 | 79/70 | 49/100 |
| **Trust in dental products score (p=0.35)** | 2.76 (1.98) | 2.89 (1.75) | 2.49 (1.86) |
| Missing | 4 | 70 | 100 |
| **Acceptance of Dentistry as a Commodity** | | | |
| **When you think of newer dental technology and what it offers (cleaning, whitening, etc.), how much would you like to have these products and services? (p=0.10)** | | | |
| I dislike a lot/I dislike/I am indifferent | 37 (26%) | 27 (34%) | 20 (42%) |
| I like | 68 (47%) | 34 (43%) | 16 (33%) |
| I like a lot | 39 (27%) | 18 (23%) | 12 (25%) |
| Missing | 5 | 70 | 101 |
| **Naturalness** | | | |
| **Would you rather have perfect (artificially improved), or less perfect but normal (natural) teeth? (p=0.01)** | | | |
| Perfect (artificially improved) | 60 (42%) | 26 (33%) | 11 (22%) |
| Less perfect (authentic) | 83 (58%) | 52 (67%) | 38 (78%) |
| Missing | 6 | 71 | 100 |
| **Character Bias** | | | |
| **If you were to see a stranger with a perfect set of healthy teeth, what would you think?** | | | |
| The person is probably vain and/or insincere/I am indifferent about their teeth **(p=0.13)** | 30 (21%) | 25 (32%) | 14 (29%) |
| I admire their nice teeth | 115 (79%) | 54 (68%) | 35 (71%) |
| Missing | 4 | 70 | 100 |

| Access to dental care and Availability of good dental care for patient and family | | | |
|---|---|---|---|
| **How much confidence do you have in being able to get good dental care for you and your family when you need it? (p=0.68)** | | | |
| Not at all/A little | 50 (34%) | 23 (30%) | 16 (33%) |
| Indifferent | 17 (12%) | 7 (9%) | 6 (12%) |
| Quite a bit | 53 (36%) | 33 (43%) | 20 (41%) |
| Very much | 26 (18%) | 14 (96%) | 7 (14%) |
| Missing | 3 | 72 | 100 |
| **How satisfied are you with your ability to find one good dentist to treat the whole family? (p=0.90)** | | | |
| Not at all | 23 (16%) | 8 (11%) | 6 (13%) |
| A little | 28 (19%) | 17 (22%) | 11 (23%) |
| Indifferent | 24 (16%) | 11 (14%) | 8 (17%) |
| Quite a bit | 44 (30%) | 29 (38%) | 17 (35%) |
| Very much | 27 (18%) | 11 (14%) | 6 (13%) |
| Missing | 3 | 73 | 101 |
| **How satisfied are you with your knowledge of where to get dental care? (p=0.37)** | | | |
| Not at all | 16 (11%) | 7 (10%) | 1 (2%) |
| A little | 38 (26%) | 15 (19%) | 16 (33%) |
| Indifferent | 13 (9%) | 8 (10%) | 6 (12%) |
| Quite a bit | 43 (29%) | 30 (39%) | 19 (39%) |
| Very much | 37 (25%) | 17 (22%) | 7 (14%) |
| Missing | 2 | 72 | 100 |
| **How satisfied are you with your ability to get dental care in an emergency? (p=0.58)** | | | |
| Not at all | 17 (11%) | 7 (9%) | 9 (18%) |
| A little | 29 (20%) | 16 (21%) | 5 (10%) |
| Indifferent | 16 (11%) | 8 (10%) | 4 (8%) |
| Quite a bit | 86 (58%) | 46 (60%) | 31 (63%) |
| Missing | 1 | 72 | 100 |
| **How satisfied are you with how convenient your dentist's office is to your home? (p=0.86)** | | | |
| A little | 33 (23%) | 18 (23%) | 13 (27%) |
| Indifferent | 26 (18%) | 9 (12%) | 5 (10%) |
| Quite a bit | 47 (32%) | 31 (40%) | 19 (39%) |
| Very much | 40 (27%) | 19 (25%) | 12 (24%) |
| Missing | 3 | 72 | 100 |
| **How difficult is it for you to get to your dentist's office. (p=0.03)** | | | |
| Not at all | 72 (50%) | 43 (56%) | 32 (65%) |
| A little | 33 (23%) | 19 (25%) | 10 (20%) |
| Indifferent | 18 (12%) | 8 (10%) | 4 (8%) |
| Quite a bit | 22 (15%) | 7 (9%) | 3 (6%) |
| Missing | 4 | 72 | 100 |

**Table 4: Summaries and longitudinal change of the outcome OHIP-14: dimensions and total/sum score**

|  | Baseline | 2 weeks - After | 4 weeks - After |
|---|---|---|---|
| **OHIP-14 Dimensions** | Median (First Quartile, Second Quartile) / (missing) | | |
| Functional | 0 [0,2] / (13) | 0 [0,2] / (71) | 0 [0,1] / (103) |
| Pain | 8 [4,12] / (23) | 4 [2,8] / (77) | 2 [0,4] / (101) |
| Psychological discomfort | 4 [2, 9] / (18) | 2 [0,5] / (72) | 2 [0,4] / (99) |
| Physical discomfort | 3 [0,6] / (15) | 2 [0,4] / (71) | 0 [0,3] / (99) |
| Psychological disability | 4 [2,8] / (15) | 2 [0,4] / (71) | 2 [0,4] / (99) |
| Sociological disability | 2 [0,6] / (8) | 0 [0,2] / (71) | 0 [0,2] / (99) |
| Handicap | 2 [0,5] / (8) | 0 [0,2] / (70) | 0 [0,3] / (99) |
| **OHIP-14 (Score)** | Mean (SD) / (missing) | | |
| **Natural-log Sum OHIP** (p=0.001) | 3.26 (.93) / (44) | 2.76 (1.03) / (87) | 2.64 (1.02) / (116) |

**Table 5: Multivariate general estimating equations models for the OHRQOL (OHIP-14)**

|  | Model 1: Gregory's Framework Covariates only | | | Model 2: Adjusted for all factors | | |
|---|---|---|---|---|---|---|
| **Outcome: log OHIP** | Coeff | 95% C.I. | P | Coeff | 95% CI | P |
| Time | -0.56 | (-0.72, -0.41) | 0.000 | -0.52 | (-0.69, -0.36) | 0.000 |
| Normative | -0.44 | (-0.73, -0.15) | 0.003 | -0.38 | (-0.65, -0.11) | 0.01 |
| Locus of Control (adherence) | 0.29 | (0.05, 0.54) | 0.02 | 0.25 | (-0.004, 0.50) | 0.05 |
| Locus of Control (external) | -0.11 | (-0.22, -0.01) | 0.04 | -0.11 | (-0.22, 0.01) | 0.07 |
| TIDP: Trust in dental products | -0.08 | (-0.14, -0.02) | 0.01 | -0.09 | (-0.16, -0.03) | 0.003 |
| Natural preference | -0.29 | (-0.57, -0.01) | 0.04 | -0.28 | (-0.57, 0.01) | 0.06 |
| Character bias | 0.32 | (0.02, 0.62) | 0.04 | 0.37 | (0.07, 0.67) | 0.02 |
| Accessibility/Availability | -0.08 | (-0.14, -0.02) | 0.01 | -0.02 | (-0.05, 0.004) | 0.10 |
| Dental Anxiety (Low Threshold) |  |  |  | 0.06 | (-0.003, 0.12) | 0.06 |
| Symptom |  |  |  | 0.29 | (-0.02, 0.60) | 0.07 |